# Independent Component Analysis for Filtering Airwaves in Seabed Logging Application

Adeel Ansari[1], Afza Bt Shafie[2], Abas B Md Said[1], Seema Ansari[3]
Electromagnetic Cluster
[1]Computer Information Sciences Department
[2]Fundamental and Applied Sciences Department
Universiti Teknologi PETRONAS
[3]Electrical Engineering Department,
Institute of Business Management, Pakistan
Adeel.ansari@hotmail.com, afza@petronas.com.my, abass@petronas.com.my, seemaansari2k3@hotmail.com

*Abstract*— Marine controlled source electromagnetic (CSEM) sensing method used for the detection of hydrocarbons based reservoirs in seabed logging application does not perform well due to the presence of the airwaves (or sea-surface). These airwaves interfere with the signal that comes from the subsurface seafloor and also tend to dominate in the receiver response at larger offsets. The task is to identify these air waves and the way they interact, and to filter them out. In this paper, a popular method for counteracting with the above stated problem scenario is Independent Component Analysis (ICA). Independent component analysis (ICA) is a statistical method for transforming an observed multidimensional or multivariate dataset into its constituent components (sources) that are statistically as independent from each other as possible. ICA-type de-convolution algorithm that is FASTICA is considered for mixed signals de-convolution and considered convenient depending upon the nature of the source and noise model. The results from the FASTICA algorithm are shown and evaluated. In this paper, we present the FASTICA algorithm for the seabed logging application.

## I. INTRODUCTION

Marine controlled source electromagnetic (CSEM) sounding that can detect and characterize offshore hydrocarbon reserves has become an important technique in oil and gas industry nowadays. This technique was introduced by [25] for offshore hydrocarbon exploration. This technique uses a horizontal electric dipole (HED) antenna as the source, emitting an alternating current typically in the range of 0.1 – 10Hz. The HED source is towed 20 – 40m above the seabed while an array of stationery EM receivers deployed on the sea floor records the resulting EM field. This technique uses the resistivity contrast as hydrocarbon reservoirs are typically known to be 5 – 100 times more resistive than host sediments [2]. Hydrocarbon reservoirs are known to have resistivity value of 30 – 500 Ωm in contrast to sea water of layer of 0.5 – 2 Ωm and sediments of 1 -2 Ωm. This high resistive reservoir will guide EM energy over long distances with low attenuation. Where highly resistive hydrocarbons are present in the subsurface, the electric fields at the receivers' at large source-receiver separations will be larger in magnitude than the more attenuated background fields passing through the host sediments [2].

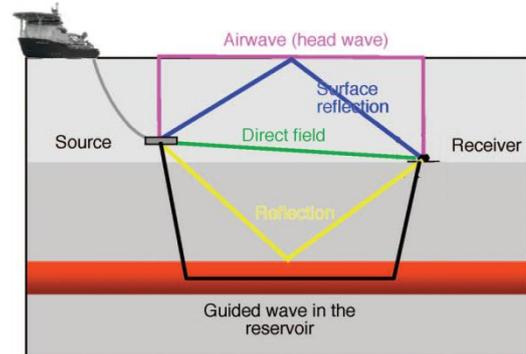

Figure 1: Simplified model used to illustrate some of the dominating field modes (events) in the EMSBL experiment.

The generated EM soundings employing this SBL technique generates three main components (waves), they are direct EM waves, guided waves (associated with high-resistivity zone like hydrocarbon reservoirs) and airwaves. The airwave component is predominantly generated by the signal component that diffuses vertically upwards (because total reflection occurs for small angles from the vertical at the sea surface) from the source to the sea surface, then propagates through the air at the speed of light with no attenuation before diffusing back down vertically through water layer to the sea bottom.

A very popular de-convolution technique considered for this domain application is the Independent Component Analysis method. ICA is a recently developed method in which the goal is to find a linear representation of non-gaussian data so that the components are statistically independent, or as independent as possible. Such a representation seems to capture the essential structure of the data in many applications, including feature extraction and signal separation. ICA is a very general-purpose statistical technique in which observed random data are linearly transformed into components that are maximally independent from each other, and simultaneously have "interesting" distributions.

ICA can be formulated as the estimation of a latent variable model. The aim for this research is to de-convolute the airwave from the signal response. A computationally very efficient method performing the actual estimation is given by the FastICA algorithm. FastICA is an efficient and popular algorithm for independent component analysis. The algorithm is based on a fixed-point iteration scheme maximizing non-Gaussianity as a measure of statistical independence. Applications of ICA can be found in many different areas such as audio processing, biomedical signal processing, image processing, telecommunications, and econometrics.

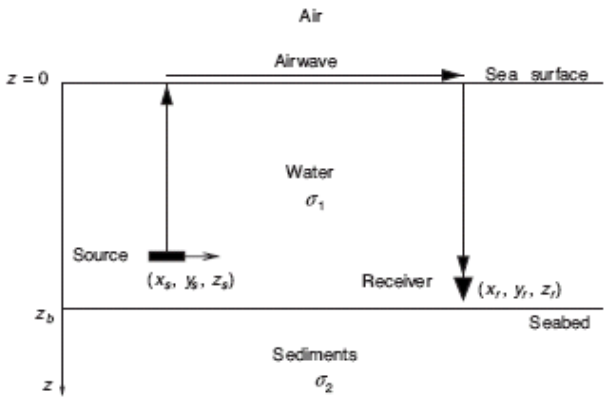

Figure 2: Airwave Generated from the Source Diffuses Up to the Air and Propagates and diffuses downward to the receivers [12].

In shallow water, this airwave component contains no information about seafloor resistivity and tends to dominate the received signal at long source receiver offsets [4]. In other words this air wave interferes with the signal that comes from the subsurface and due to this its present is considered as unwanted signal [12].

Using Computer Simulation Technology (CST) software we will set up a simulation model that contains air, sea water, sediments and hydrocarbon reservoir. The parametric settings of the sea water depth is varied from deep water (2000m) to shallow water (100m).

Independent Component Analysis has been considered as a popular method for mixed signal de-convolution and has been progressing with the advent of time on other problem settings. The extension presented in this research study is by far not the only possible ones. Suitable ICA-type algorithms like FASTICA, Infomax and PCA Orthogonal mixing are proposed with respect to the source and noise model, for the identification and filtration of the airwaves in seabed logging application.

## II. PROBLEM STATEMENT

- The presence of airwaves in the signal response effects the detection of Hydrocarbon presence in the seabed logging application.
- A technique is required for filtering the airwaves from the signal response.
- Selection of appropriate algorithm is required for filtering the airwaves.
- The algorithm must filter out the airwave component from the loaded data.

## III. LITERATURE REVIEW

Independent component analysis was originally developed to deal with problems that are closely related to the cocktail-party problem. Since the recent increase of interest in ICA, it has become clear that this principle has a lot of other interesting applications as well.

Another, very different application of ICA is on feature extraction. A fundamental problem in digital signal processing is to find suitable representations for image, audio or other kind of data for tasks like compression and denoising.

All of the applications described above can actually be formulated in a unified mathematical framework, that of ICA. This is a very general-purpose method of signal processing and data analysis.

In this review, we cover the definition and underlying principles of ICA. The domain application for ICA in this research is over Seabed Logging and the airwave component which is also illustrated in this review.

### A. Independent Component Analysis

Independent Component Analysis (ICA) is an important tool for modeling and understanding empirical datasets as it offers an elegant and practical methodology for blind source separation and deconvolution. It is now possible to observe a pure unadulterated signal from a mixture of signals usually corrupted by noise via statistical approach.
As the field of signal processing is greatly concerned with the problem of recovering the constituent sources from the convolutive mixture; ICA maybe applied to this Airwave Source Separation problem to recover the sources.

### 1) FASTICA Algorithm

The algorithm is based on a fixed-point iteration scheme maximizing non-Gaussianity as a measure of statistical independence.
The Electric field intensity of the Ex component obtained from CST simulation is considered as different data samples and arranged as a single column matrix **X**. The results from the simulated airwave is considered as the second column of the matrix X, so as to obtain the two number of components, one matching with the airwave results.

$$X = As \ldots \ldots \ldots (1)$$

Where $s = s_1 + s_{air}$

We know $\mathbf{s_{air}}$, we need to find $\mathbf{s_1}$, which contains no airwave.

$$S = Wx \ldots \ldots \ldots (2)$$

Where $W = A^{-1}$,

Kurtosis technique is suitable to measure the non-gausianity as the E-field intensity is more Gaussian in nature and the E-field strength is the measured data sample to ascertain the unmixed sources. The algorithm works iteratively for each individual ICs, that is, two.

Data Pre-processing:
Since the E-field strength values are very small, we have assigned weights to each strength value for offset from 0 to 25,000m, so as to perform the FASTICA algorithm efficiently.

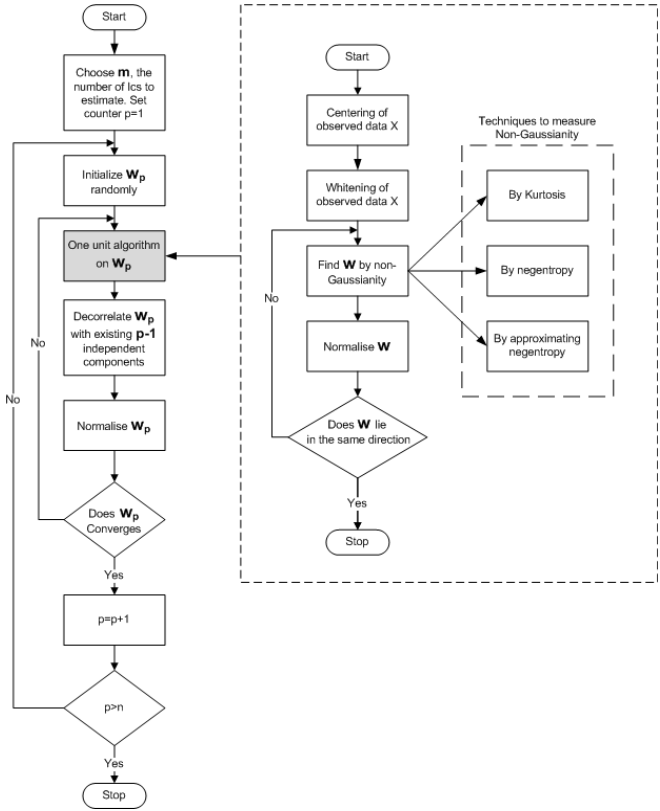

Figure 3: Flowchart of FASTICA Algorithm.

The FastICA algorithm and the underlying contrast functions have a number of desirable properties when compared with existing methods for ICA.

1. The convergence is cubic (or at least quadratic), under the assumption of the ICA data model. This is in contrast to ordinary ICA algorithms based on (stochastic) gradient descent methods, where the convergence is only linear. This means a very fast convergence, as has been confirmed by simulations and experiments on real.
2. Contrary to gradient-based algorithms, there are no step size parameters to choose. This means that the algorithm is easy to use.
3. The algorithm finds directly independent components of (practically) any non-Gaussian distribution using any nonlinearity. This is in contrast to many algorithms, where some estimate of the probability distribution function has to be first available, and the nonlinearity must be chosen accordingly.
4. The performance of the method can be optimized by choosing a suitable nonlinearity. In particular, one can obtain algorithms that are robust and/or of minimum variance.
5. The independent components can be estimated one by one, which is roughly equivalent to doing projection pursuit. This is useful in exploratory data analysis, and decreases the computational load of the method in cases where only some of the independent components need to be estimated.
6. The FastICA has most of the advantages of neural algorithms: It is parallel, distributed, computationally simple, and requires little memory space. Stochastic gradient methods seem to be preferable only if fast adaptivity in a changing environment is required.

*2) Assumptions:*

1. Sources are considered as independent and linearly mixed by a mixing matrix $\mathbf{A}$.
2. We cannot determine the variances (energy levels) of each independent component.
3. We cannot determine the order of the independent components.
4. $\mathbf{s}$ and $\mathbf{n}$ have mean zero and consequently $\mathbf{x}$ has zero mean.

### B. The Sea bed Logging Method

Sea bed logging utilizes controlled source electromagnetic (EM) sounding technique in detecting subsurface hydrocarbon. CSEM method employs a horizontal electric dipole (HED) source to transmit low frequency (typically 0.01 – 10Hz) signals to an array of receivers that measure the electromagnetic field at the seafloor. By studying the variation in amplitude and phase of the received signal as the source is being towed over the receiver array, the resistivity structure of the subsurface can be determined at a depth of several kilometers [6]. According to [3], as depicted by Figure 3, the receivers record the EM responses as a combination of energy pathways including signal transmitted directly through seawater, reflection and refraction via the sea-water interface, refraction and reflection along the sea bed and reflection and refraction via possible high resistivity subsurface layers.

Electromagnetic waves attenuate rapidly in seawater and sediments, and they tend to dominate in greater energy at very closer offsets. Hydrocarbon filled reservoirs have a higher resistivity of 30 - 500Ωm and the electromagnetic waves are easily guided along the layers and attenuate less depending on the critical angle of incidence. Guided EM energy is constantly refracted back to the seafloor and recorded by the EM receivers. This source energy is also reflected and refracted via air-water interface and is called an air wave and it starts dominating at far offsets (about 6km).

The seawater depth has a strong influence on the measured signal response due to the presence of air wave effect on shallow water environment. This airwave effect is very much eminent within shallow water environment at greater intensity, due to which the useful signals from the reservoir or targets get totally masked by the airwaves which contain little information about subsurface [3]. Due to this, we investigate the air wave effect as we increase water level gradually and the filtration technique ICA that is proposed for this airwave filtration.

## C. The Airwave

Airwaves components are energies that diffuse from the source within the seawater and travel vertically upwards towards the surface and propagate at the speed of light with no attenuation before diffusing back down vertically through water layer to the receiver on the seabed. Airwave component is problematic in shallow water because it is less attenuated during its up and down propagation due to the short sea water depth.

According to [1] the airwave is guided at the air-water interface with a decay of approximately $1/r^3$, where $r$ is the offset, for a far offsets of source and receivers. For shallow sea water depth, the air waves are stronger and mask the signal from the hydrocarbon and it is considered as noise.

The air wave expression at offset $r$ in the radial direction with z-axis in positive downwards direction with $s_z$ and $r_z$ denoting the depth of source and receivers respectively. The asymptotic space domain expression for the air wave is given by

$$E_r^{(air)} = \frac{p\cos\emptyset \exp[ik(z_r+z_s)]\exp[ik_0 r]}{2\pi\sigma_r r^3} \quad (2)$$

where $p$ is the HED source dipole moment, $\emptyset$ is the azimuth angle of the source,

$$k = (i\omega\mu_0\sigma_1)^{1/2} = (1+i)k$$

$k$ is the complex low frequency wave number for sea water with conductivity $\sigma_1$,

$$k^{-1} = (\omega\mu_0\sigma_1)^{1/2}$$

$k^{-1}$ is the skin depth in sea water,

$k_0 = \omega(\mu_0\varepsilon_0)^{1/2} \approx 0 m^{-1}$

$k_0$ is the wave number in air, $\omega$ is the circular frequency, $\mu_0$ is the magnetic permeability in vacuum and $\varepsilon_0$ is the permittivity in vacuum. This equation is used by [4] to demonstrate the behavior of the air wave component in marine CSEM surveying [12].

## IV. RESEARCH OBJECTIVES

- To identify, characterize and filter the air waves within the signal response received from the receiver.
- Perform simulations by varying the parameters of the Hydrocarbon reservoir, so as to analyze the behavior of the airwaves within the signal response at the receiver end, within seabed logging application.
- To identify and apply the ICA statistical technique for filtering the airwaves from the signal response.
- To carry out a performance-based evaluation and validation of the appropriate ICA algorithm suitable for filtering the airwaves.

## V. METHODOLOGY

The methodology and approach for this research study will be conducted in five stages as shown in the Figure 8.

The first stage is very crucial and involves most of the literature review and related work portion.

In order to decipher the airwaves from the signal response, we must first obtain thorough knowledge of the seabed logging application and about the electromagnetic waves properties at low frequencies under seawater and in the oceanic lithosphere. As said before, this stage is crucial, without which will become difficult to identify and characterize the airwaves required for filtration using ICA technique, which is our primary objective, to identify and filter out the airwaves in seabed logging.

The second stage will comprise of the simulation work that is required to:

- Understand the variations of the electric field strength at varying offsets by varying the properties of the simulation model.
- The simulation working will allow us to understand the presence of other wave components (airwave inclusively) within the signal response by varying the distance between the transmitter and the receiver at various seawater depths.

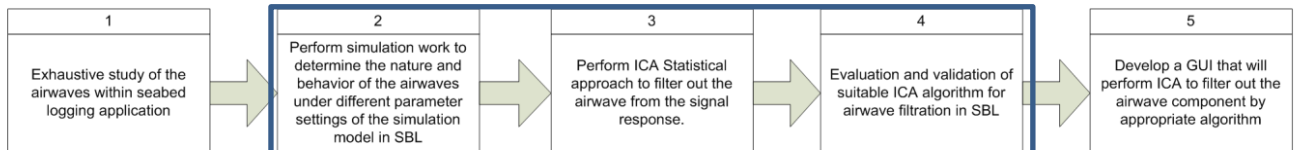

Figure 4: Methodological approach for ICA statistical approach to airwave filtration.

- To singly determine the presence of the airwaves present within the signal response

The third stage is segregated into two folds. The first is the Linear Mixing fold and the second fold pertains to implying of the ICA statistical approach to filter out the airwaves from the obtained signal response.

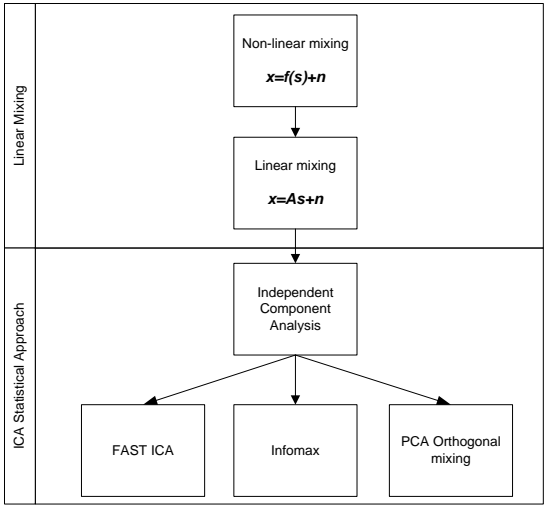

Figure 5: Stage 3 is divided into two folds, the Linear Mixing fold and the ICA Statistical Approach fold.

## VI. RESULTS AND DISCUSSIONS

The simulation model as shown in Figure 7, has been considered primary to work with the FASTICA algorithm. The conductivity model for each medium layer used in the model (see Figure 8).

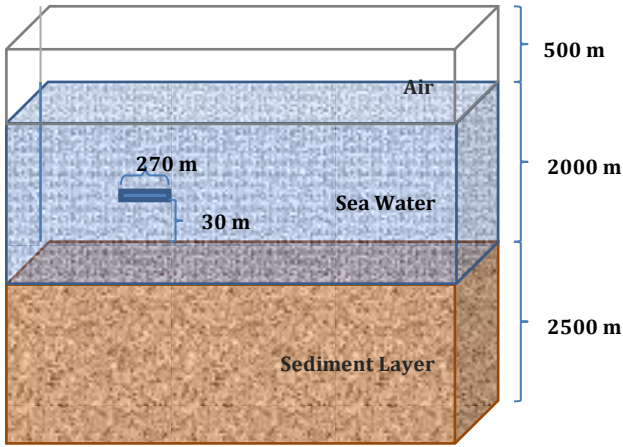

Figure 6. Simulation Model 1 showing no presence of Hydrocarbon reservoir.

Figure 7: Horizontally layered configuration. The source at depth of 30m from seafloor operates at 0.125 Hz. The receivers are located at the sea bottom, with a total extent of 50km in the x & z direction.

In this research progress, various FASTICA results have been obtained for weights $10^5, 10^6$ and $10^7$ and the graphs are then compared with the graph results from the air removal method as mentioned in [5]. The results obtained are for seawater depth of 100m, 300m, 500m, 700m, 900m and 1000m. The airwave was generated through numerical modeling from equation 1 in [2] and has been calculated for various seawater depth.

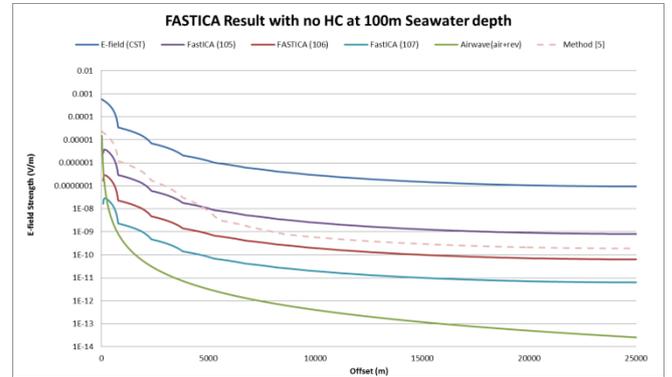

Figure 8. Comparison of airwave with the Electric field intensity for seawater depth of 100m with FASTICA results of varying weights used also in comparison is air removal method [5].

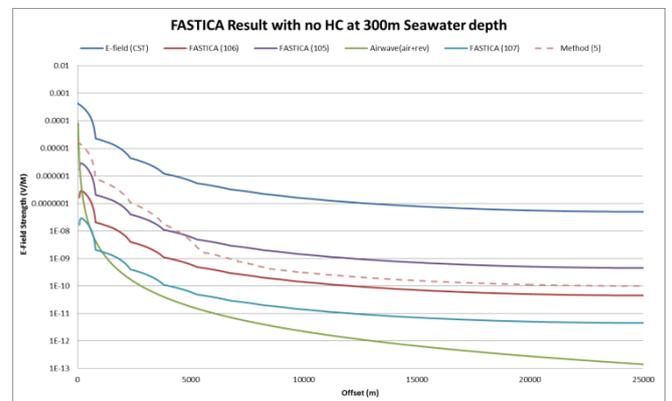

Figure 9. Comparison of airwave with the Electric field intensity for seawater depth of 300m with FASTICA results of varying weights used also in comparison is air removal method [5].

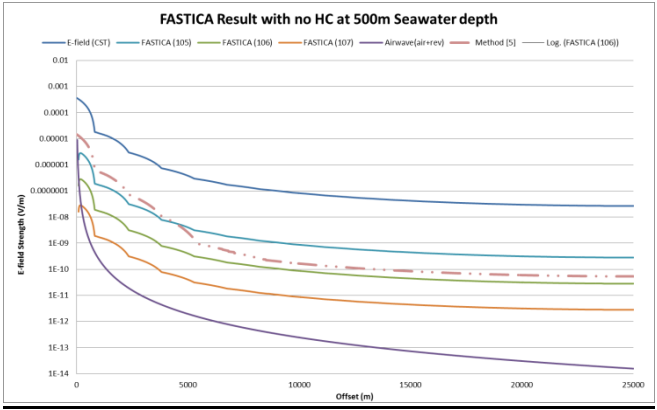

Figure 10. Comparison of airwave with the Electric field intensity for seawater depth of 500m with FASTICA results of varying weights used also in comparison is air removal method [5].

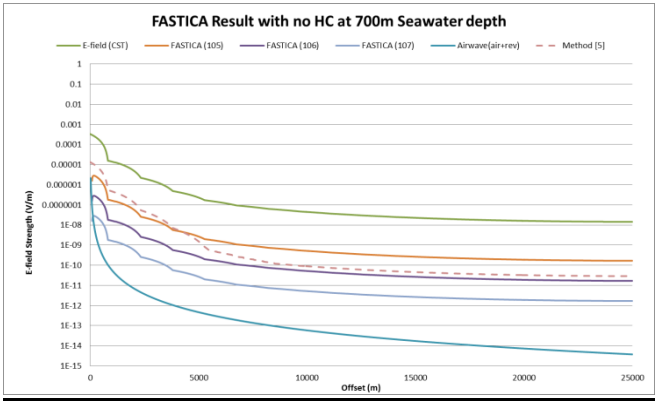

Figure 11. Comparison of airwave with the Electric field intensity for seawater depth of 700m with FASTICA results of varying weights used also in comparison is air removal method [5].

The reason for selecting weights of range from $10^5$ till $10^7$ is because the sample values of the E-field intensity is of very minute values ranging from $10^6$ till $10^{15}$.

From Table I and from Figure 8 to 13, it is very evident that the magnitude of the FASTICA result of weight $10^6$ is very close to the results from [5] for far offset from 15km till 25 km. In shallow water setting, the EM waves reverberate more often, resulting in higher electric field strength at the receiver response.

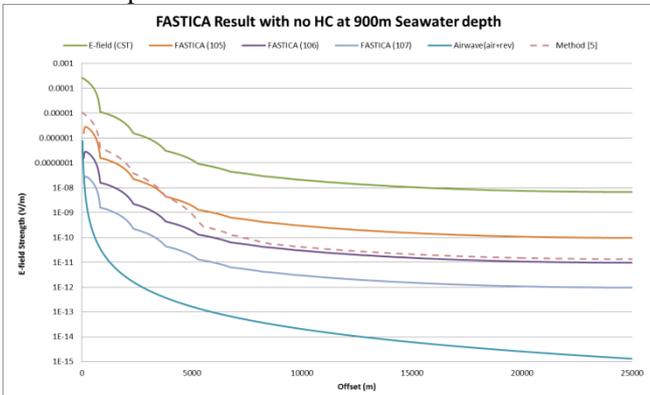

Figure 12. Comparison of airwave with the Electric field intensity for seawater depth of 100m with FASTICA results of varying weights used also in comparison is air removal method [5].

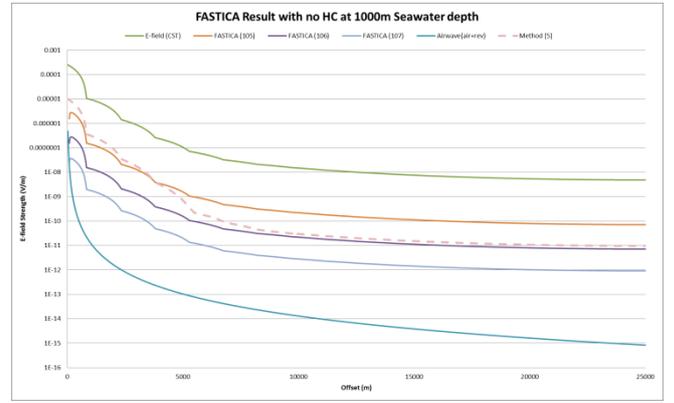

Figure 13. Comparison of airwave with the Electric field intensity for seawater depth of 100m with FASTICA results of varying weights used also in comparison is air removal method [5].

Table I. Comparison of magnitude of FASTICA results for varying seawater depth at Offset 25,000m of varying weights with Method of air removal from [5].

| Seawater Depth (m) | Results from FASTICA ($10^5$) | Results from FASTICA ($10^6$) | Results from FASTICA ($10^7$) |
|---|---|---|---|
| 1000 | 643% | -26% | -91% |
| 900 | 618% | -28% | -93% |
| 700 | 482% | -42% | -94% |
| 500 | 426% | -47% | -95% |
| 300 | 353% | -55% | -95% |
| 100 | 330% | -66% | -97% |

between the CST simulation recordings with the calculated airwave.

This is the main reason as to why the airwaves mask the important informative signals at far offsets. The FASTICA results for each seawater depth are at an intermittent level

The graphs of FASTICA results of varying weights $10^5$, $10^6$ and $10^7$ are equidistant from each other, with the $10^5$ weight having most E-field strength, whilst the weight $10^7$, the E-field strength is significantly reduced, resulting in a graph slightly closer to the actual airwave.

The airwave removal results from [5] are projected and compared with the FASTICA results of varying weights so as to ascertain the reliability in the results obtained from FASTICA. Since the results from [5] coincide significantly with the FASTICA result of weight $10^6$ by a percentage difference of minimum 26%, also we consider the offset from 15km till 25km as airwaves are more imminent at a far off distance. Hence from conclusion, we can say that the FASTICA result of weight $10^6$ gives better results of the EM response without the airwaves.

## VII. CONCLUSION

In conclusion, we have performed FASTICA algorithm over the CST simulation results based over the no Hydrocarbon simulation model by varying the weights used. The results from the FASTICA of weights $10^5$, $10^6$ and $10^7$ are determined and compared with the air wave removal method in [5] for validation of the results. The weight $10^6$ coincides with the results from [5] at far offsets from 15km till 25km. Hence weight $10^6$ approximate the best results.

Since we have now obtained the result set from FASTICA algorithm in deciphering out the airwaves, future work is to commence with the other algorithms that is, PCA-orthogonal mixing algorithm using ICA. Along with that, we will use complex simulation modeling environments for algorithm testing for evaluation and validation.